\documentclass[11pt]{article}
\usepackage{amsfonts}
\usepackage[english]{babel}
\usepackage{graphicx}
\usepackage{epstopdf}
\usepackage{epsfig}
\usepackage{amssymb}
\usepackage{setspace}
\usepackage{caption}
\usepackage{amsfonts}
\usepackage{color}
\usepackage{amsmath}
\usepackage{float}
\usepackage{comment}
\newcommand {\be}{\begin{equation}}
\newcommand {\ee}{\end{equation}}
\newcommand {\bea}{\begin{array}}
	
	\newcommand {\eea}{\end{array}}
\newcommand{\sch}{Schwarzschild~}

\begin{document}

	\begin{titlepage}
		\vspace{1cm} 
		\begin{center}
			{\Large \bf {Merger estimates for Kerr-Sen black holes}}\\
		\end{center}
		\vspace{2cm}
		\begin{center}
			\renewcommand{\thefootnote}{\fnsymbol{footnote}}
			Haryanto M. Siahaan{\footnote{haryanto.siahaan@unpar.ac.id}}\\
			Center for Theoretical Physics,\\ Department of Physics, Parahyangan Catholic University,\\
			Jalan Ciumbuleuit 94, Bandung 40141, Indonesia
			\renewcommand{\thefootnote}{\arabic{footnote}}
\end{center}
		
\begin{abstract}
The advent of gravitational wave observation starts a new era of precision tests of gravitational theory. Estimations of black holes mergers should come from any well established gravitational theories, provided that the theory has not been ruled out by observations. In this paper we consider the low energy limit of heterotic string theory where the associated rotating and charged black holes are described by the Kerr-Sen solution. We investigate the approximate final spins and quasinormal modes of the black hole resulting from the merger process.
\end{abstract}
	\end{titlepage}\onecolumn 
	\bigskip 
	
	\section{Introduction}
	\label{sec:intro}
	
The remarkable gravitational wave (GW) observations of GW150914 \cite{Abbott:2016blz}, GW151226 \cite{Abbott:2016nmj}, and GW170104 \cite{Abbott:2017vtc} are the beginning of new era in astronomy. Instead of observation that relies on electromagnetic wave only, now astronomers can also see the sky using GWs. Moreover, some information on black holes properties can be extracted from GWs produced by black hole mergers. As we know, black hole is an object that requires our deep understanding on gravity in strong regime. Hence, observation of GWs produced by black hole collisions may put the ordinary Einstein theory of gravity into a rigid test. 
	
However, general relativity (GR) appears still profound so far in predicting the observed GW produced by binary black holes merger \cite{Yunes:2016jcc}. Suppose that there exist small deviations to the GR predictions, revealing them would be another difficulties to be dealt with due to noises in the signals. Nevertheless, while the improvement on extracting data method from GW is in progress, possible deviations on black holes merger estimates coming from any well established gravitational theory is worth to be investigated. Nevertheless, the fact that numerical study for this estimations is very costly suggests that the early investigations on black hole mergers should make use of some approximate methods. Examples of these approaches are the Buonanno-Kidder-Lehner (BKL) recipe to estimate the final spins after merger \cite{Buonanno:2007sv} and approaching QNMs frequency of final black hole using light ring properties \cite{Cardoso:2008bp}. Since these are just some approximations, several limitations to the proposal in \cite{Cardoso:2008bp} have been reported \cite{Khanna:2016yow,Konoplya:2017wot}.

In this paper we consider the Kerr-Sen black holes of the low energy limit heterotic string theory \cite{Sen:1992ua}. We would like to study the estimates of black holes mergers which result a Kerr-Sen black hole in the line with the work presented in \cite{Jai-akson:2017ldo} where Kerr-Newman black hole is studied. Kerr-Sen and Kerr-Newman black holes are analogous, but not exactly the same \cite{Ghezelbash:2012qn}. The two black hole solutions are asymptotically flat, rotating, and electrically charged. The neutral limit of the two solutions are also the same, namely the Kerr solution. Moreover, both solutions can be equipped with Taub-NUT parameter which then yields the notion of black hole becomes obscure due to the existence of conical singularity \cite{Griffiths:2009dfa,Siahaan:2019kbw}. 

The nature that Kerr-Sen solution belongs to the low energy limit of string theory yields the solution is worth for further studies. It is because string theory is the present strongest candidate for quantum description of gravity. The same motivation has also inspired some recent works to discuss aspects of black holes in the low energy limit of string theory \cite{Duztas:2018adf,Duztas:2018ebr,Vieira:2018hij,Gwak:2016gwj,Uniyal:2017yll,Liu:2018vea,Siahaan:2015ljs,Bernard:2017rcw}. It is one of our aims to search some possible different features between the Kerr-Sen and Kerr-Newman black holes from the merger estimates point of view. 

In particular, the estimates that we would like to study are the spin and quasinormal modes (QNMs) of the final black hole. In making the final spin estimations, we employ the generalized BKL recipe by Jai-akson et al \cite{Jai-akson:2017ldo} which applies to rotating and charged black holes. In fact, the work presented in this paper is motivated by the results reported in \cite{Jai-akson:2017ldo} where the mergers of Kerr-Newman and Kaluza-Klein black holes in Einstein-Maxwell-(dilaton) theory are investigated. In addition to the dilaton and gauge fields in the Einstein-Maxwell-dilaton theory, the low energy limit of heterotic string field action consists of an antisymmetric second rank tensor field. Despite is is very unlikely for a collapsing objects to maintain a significant electric charge, studies related to charged black holes coallesence exist in literature \cite{Liebling:2016orx,Fraschetti:2016bpm,Zhang:2016rli,Zilhao:2012gp}. Surely the electric charge can potentially contribute to a considerable difference of black holes merger estimations compared to the case where electric charge is absent. 

To approximate the QNMs we follow the prescription by Cardoso et al \cite{Cardoso:2008bp} where the light ring geodesic is the main ingredient. According this approach, limited in the eikonal limit regime, the real part of QNM is related to the angular velocity of null objects orbiting around the light ring. On the other hand, the imaginary part is considered as the Lyapunov exponent of the unstable light ring. Despite there exist some gaps between the QNM numerical results and the ones given by the light ring approach, the deviation is still acceptable to be used as a guidance to do further numerical work on predictions made by alternative theories of gravity. 
	
The organization of this paper is as follows. In section \ref{s2} we present short a review on the low energy limit of string theory and Kerr-Sen black hole. In the next section, we provide some generalities including the circular geodesics and the generalized BKL method. In section \ref{s4}, we study the merger estimations made in pure geodesic consideration, i.e. the timelike orbiting probe is neutral. In the next section, we turn our discussion to the charged orbiting probe, where corrections coming from the Coulomb interaction between the objects appear. In section \ref{s6}, some comparisons of final spins between Kerr-Newman and Kerr-Sen mergers are given. Finally, we give our comments and conclusions in the last section. 

\section{Low energy heterotic string theory black holes}\label{s2}
	
An effective action describing fields in low energy of heterotic string theory can be read as
\be\label{action.het} S = \int {{d^4}x\sqrt { - g} ~{\exp \left( { - \Phi } \right)}\left( {R + {g^{\mu \nu }}{\partial _\mu }{\partial _\nu }\Phi  - \frac{1}{8}{g^{\alpha \mu }}{g^{\beta \nu }}{F_{\alpha \beta }}{F_{\mu \nu }} - \frac{1}{{12}}{g^{\alpha \mu }}{g^{\beta \nu }}{g^{\chi \kappa }}{H_{\alpha \beta \chi }}{H_{\mu \nu \kappa }}} \right)} \ee 
	where ${F_{\mu \nu }} = {\partial _\mu }{A_\nu } - {\partial _\nu }{A_\mu }$ and ${H_{\alpha \beta \chi }} = {\partial _\alpha }{B_{\beta \chi }} + {\partial _\chi }{B_{\alpha \beta }} + {\partial _\beta }{B_{\chi \alpha }} - \frac{1}{4}\left( {{A_\alpha }{F_{\beta \chi }} + {A_\chi }{F_{\alpha \beta }} + {A_\beta }{F_{\chi \alpha }}} \right)$.
Fields incorporated in this theory are the spacetime metric $g_{\mu\nu}$, $U\left(1\right)$ vector field $A_\mu$, dilaton field $\Phi$, and the second-rank antisymmetric tensor field $B_{\mu\nu}$. The following is a set of field solutions for the classical equations of motion obtained from (\ref{action.het}) as reported in \cite{Sen:1992ua}. The spacetime metric expressed in Einstein frame reads 
	\[d{s^2} =  - \left( {1 - \frac{{2Mr}}{{{\rho ^2}}}} \right){\rm{d}}{t^2} - \frac{{4Mra{{\left( {1 - x^2 } \right)}}{\rm{d}}t{\rm{d}}\phi}}{{{\rho ^2}}} + {\rho ^2}\left( {\frac{{{\rm{d}}{r^2}}}{{{\Delta }}} + \frac{{{\rm{d}}{x^2}}}{{{\left( {1 - x^2 } \right)}}}} \right)\]
	\be\label{KerrSenmetricEinsteinFrame} + \left( {{\rho ^2} + {a^2}\left(1-{x^2}\right) + \frac{{2Mr{a^2}{\left( {1 - x^2 } \right)}}}{{{\rho ^2}}}} \right){\left( {1 - x^2 } \right)}{\rm{d}}{\varphi ^2}\,,\ee
	where ${\Delta } = r\left( {r + 2b} \right) - 2Mr + {a^2}$, and ${\rho ^2} = r\left( {r + 2b} \right) + {a^2}{x^2}$. The accompanying non-gravitational fields solutions are
	\be \Phi  =  - \frac{1}{2}\ln \frac{{{\rho ^2}}}{{{r^2} + {a^2}{x^2}}}\,,\ee
	\be\label{KerrSenVectors} {A_\mu }{\rm{d}}{x^\mu } = \frac{{Qr}}{{{\rho ^2}}}\left( {{\rm{d}}t - a{\Delta _x}{\rm{d}}\varphi } \right)\,,\ee
	and
	\be\label{2ndrankTensor} {B_{t\varphi }} =  - {B_{\varphi t}} = \frac{{bra{\Delta _x}}}{{{\rho ^2}}}\,.\ee   

The line element (\ref{KerrSenmetricEinsteinFrame}) is asymptotically flat and contains a black hole solution namely the Kerr-Sen black hole. The black hole mass is $M$, rotational parameter is $a = J/M$, and the electric charge is $Q = \sqrt {2bM} $. The mass and angular momentum can be computed using Komar integral, using the timelike and axial Killing vectors. Just like Kerr-Newman case, the non-extremal Kerr-Sen black holes have two separate horizons namely the outer and inner horizons, $r_+$ and $r_-$ respectively. These horizon locations are the root of $\Delta$ which are $r_ \pm   = M - b \pm \sqrt {\left( {M - b} \right)^2  - a^2 }$. The extremality of Kerr-Sen black holes occurs at $M = a+b$ which yields $r_+ = r_-$. The area of Kerr-Sen black hole is given by $A_{BH} = 8\pi Mr_ +$, and accordingly the entropy is 
	\be 
	S_{BH} = \frac{A_{BH}}{4} = 2\pi Mr_+\,.
	\ee 
	The non-rotating limit of Kerr-Sen black hole, which yields the vanishing of $B_{\mu\nu}$, is the Gibbons-Maeda-Garfinkle-Horowitz-Strominger (GMGHS) black hole \cite{GMGHS}. This solution is analogous to the Reissner- Nordstrom black hole, but again not exactly the same. The Kerr-Sen solution (\ref{KerrSenmetricEinsteinFrame}) - (\ref{2ndrankTensor}) has been generalized to the accelerating objects \cite{Siahaan:2018qcw} and the spacetime containing Taub-NUT parameter \cite{Siahaan:2019kbw}.

	\section{Generalities}\label{s3}
	
	\subsection{Timelike and null geodesics}
	
	The corresponding Lagrangian for a charged massive test particle is
	\be \label{LagrangianTEST}
	{\cal L} = \frac{m}{2}g_{\mu \nu } \dot x^\mu  \dot x^\nu   - qA_\mu  \dot x^\mu  \,.
	\ee
	In an axial symmetric and stationary spacetime, one can obtain the conserved quantities related to an object described by the Lagrangian (\ref{LagrangianTEST}) which read
	\be\label{Egen}
	\frac{{\partial {\cal L}}}{{d\dot t}} =  g_{tt} \dot t + g_{t\phi } \dot \phi  - e A_t \equiv  - E \,,
	\ee
	\be\label{Lgen}
	\frac{{\partial {\cal L}}}{{d\dot \phi}} = g_{t\phi } \dot t + g_{\phi \phi } \dot \phi  - e A_\phi  \equiv L \,.
	\ee
Above, $e \equiv m^{-1} q$ is the charge to mass ratio of the test object or probe. Accordingly, the last two equations give us
	\be
	\dot t = \frac{{g_{t\phi } \left( {L + eA_\phi  } \right) + g_{\phi \phi } \left( {E - eA_t } \right)}}{\Delta }\,,
	\ee
	\be
	- \dot \phi  = \frac{{g_{tt} \left( {L + eA_\phi  } \right) + g_{t\phi } \left( {E - eA_t } \right)}}{\Delta }\,.
	\ee
	
In eq. (\ref{KerrSenmetricEinsteinFrame}), we use mostly positive type of the metric. Therefore, the geodesics in general can be expressed as
	\be \label{eqGEO}
	g_{tt} \dot t^2  + g_{rr} \dot r^2  + 2g_{t\phi } \dot t\dot \phi  + g_{\phi \phi } \dot \phi ^2  =  - \delta\,,
	\ee
where $\delta = 1$ represents timelike particle and $\delta = 0$ belongs to null object. On equatorial plane, where the identity $\Delta  = g_{t\phi }^2  - g_{tt} g_{\phi \phi } $ applies, and the timelike effective potential $V_{{\rm{eff}}} \left( r \right) = \dot r^2 $ can be written as
	\be \label{VeffTIMELIKE}
	V_{{\rm{eff}}}  = \frac{{g_{tt} \left( {L + eA_\phi  } \right)^2  + 2g_{t\phi } \left( {L + eA_\phi  } \right)\left( {E - eA_t } \right) + g_{\phi \phi } \left( {E - eA_t } \right)^2  - \Delta }}{{g_{rr} \Delta }}\,.
	\ee 
The constants of motion $E$ and $L$ for an object with circular motion $\dot r = 0$ can be obtained by solving $V_{\rm eff} = 0$ and $V'_{\rm eff} = 0$ simultaneously. Explicitly, these two equations read
	\be \label{eqVgen}
	g_{tt} \left( {L + eA_\phi  } \right)^2  + 2g_{t\phi } \left( {L + eA_\phi  } \right)\left( {E - eA_t } \right) + g_{\phi \phi } \left( {E - eA_t } \right)^2  - \Delta  = 0\,,
	\ee 
	and
	\[
	\left( {E - eA_t } \right)^2 \left( {g'_{\phi \phi }  + \frac{{2eA'_t g_{\phi \phi } }}{{\left( {E - eA_t } \right)}}} \right) + \left( {L + eA_\phi  } \right)^2 \left( {g'_{tt}  + \frac{{2eA'_\phi  g_{tt} }}{{\left( {L + eA_\phi  } \right)}}} \right) - 2\left\{ {\left( {E - eA_t } \right)\left( {L + eA_\phi  } \right)g'_{t\phi } } \right.
	\]
	\be \label{eqdVgen}
	\left. { + e^2 \left( {A_t A_\phi  } \right)' - EeA'_\phi   + LeA'_t } \right\} - \Delta ' = 0 \,.
	\ee
The vanishing of $V''_{{\rm{eff}}}$ from which one can obtain the ISCO radius has an explicit form
	\[
	- \Delta '' + \left( {L + eA_\phi  } \right)^2 g''_{tt}  + \left( {E - eA_t } \right)^2 g''_{\phi \phi }  + 2\left( {E - eA_t } \right)\left( {L + eA_\phi  } \right)g''_{t\phi } 
	\]
	\[
	+ 2e\left( {L + eA_\phi  } \right)\left( {g_{tt} A'_\phi   - g_{t\phi } A'_t  + 2A'_\phi  g'_{tt}  - 2A'_t g'_{t\phi } } \right)
	\]
	\[
	+ 2e\left( {E - eA_t } \right)\left( {g_{t\phi } A'_\phi   - g_{\phi \phi } A'_t  + 2A'_\phi  g'_{t\phi }  - 2A'_t g'_{\phi \phi } } \right)
	\]
	\be \label{eqddVgen}
	+ 2e^2 \left( {g_{tt} \left( {A'_\phi  } \right)^2  + g_{\phi \phi } \left( {A'_t } \right)^2  - 2g_{t\phi } A'_\phi  A'_t } \right) = 0\,.
	\ee 
Before solving the last equation, one needs to plug the constants of motion $E$ and $L$ from (\ref{eqVgen}) and (\ref{eqdVgen}) into that equation. Simply put, the three equations $V_{{\rm{eff}}}=0$, $V'_{{\rm{eff}}}=0$, and $V''_{{\rm{eff}}}=0$ give the information $\left\{ {E,L,r_{ISCO} } \right\}$ of a circularly moving timelike object. 
	
Normally, solving the three equations (\ref{eqVgen}), (\ref{eqdVgen}), and (\ref{eqddVgen}) analytically to get the explicit expressions of $E$, $L$, and $r_{\rm ISCO}$ in the geometry of rotating and charged black holes are not so easy. Moreover, even if these three equations can be analytically solved to express exact forms of $E$, $L$, and $r_{\rm ISCO}$, it is not too straightforward to extract the corresponding qualitative implications due to their complicated forms. That is why evaluating (\ref{eqVgen}), (\ref{eqdVgen}), and (\ref{eqddVgen}) numerically is sometime the best option that we can do to study the behavior of a test object under consideration. The obtained plots could be the angular momentum of a test object vs. radius, ISCO radius vs. black hole mass, or something else depending on what we need to know. In getting the plot, some of the physical parameters need to be fixed so the result cannot be general. Clearly, the family of physical parameters of a rotating and charged black holes is larger compared to the neutral one, and therefore the parameter for $L$ or ISCO radius is no longer just black the hole rotation. They can also be the total black hole charge $Q$ or even the ratio of black hole charges before collision. 
	
As we have mentioned earlier, one of our goals is to obtain the QNM frequencies of the final black hole using the light ring approximation \cite{Cardoso:2008bp}. This requires the effective potential for null geodesics, where the correction of electric charge can come from the final spin which incorporate the charge of black holes. In null geodesic consideration, the conserved quantities $E$ and $L$ are those in (\ref{Egen}) and (\ref{Lgen}), while the geodesic equation is (\ref{eqGEO}) with $\delta =0$. This effective potential reads
	\be \label{VeffLightRing}
	V_{{\rm{eff}}}  =  - \frac{{g_{tt} \dot t^2  + 2g_{t\phi } \dot t\dot \phi  +  + g_{\phi \phi } \dot \phi ^2 }}{{g_{rr} }}\,,
	\ee 
which can be rewritten as
\be\label{Veff.null}
V_{{\rm{eff}}}  = \frac{{g_{tt} L^2  + 2g_{t\phi } LE + g_{\phi \phi } E^2 }}{{g_{rr} \Delta }}\,.
\ee 
The circular null orbits which are of our interests are dictated by $V_{{\rm{eff}}}  = 0$ and $V'_{{\rm{eff}}}  = 0$, which explicitly can be expressed as
	\be \label{pure.eqV}
	g_{tt} L^2 + 2g_{t\phi } LE  + g_{\phi \phi } E^2  = 0\,,
	\ee 
	\be \label{pure.eqdV}
	g'_{tt} L^2  + 2g'_{t\phi } LE  + g'_{\phi \phi } E^2  = 0\,.
	\ee 
The last two equations can give us the null radius of circular trajectory $r_{\rm c}$, where the unstable $r_{\rm c}$ occurs when $V''_{{\rm{eff}}} \left( {r_c } \right) > 0$.

Following \cite{Cardoso:2008bp}, the QNM frequencies $\omega_{QNM}$ can be approached by the light ring geodesic properties. Using the angular velocity
	\be \label{OmegaCgen}
\Omega _c  = \left. {\frac{{\dot \phi }}{{\dot t}}} \right|_{r_c }  \,,
\ee
and the Lyapunov exponent 
	\be\label{lambdaGEN} 
\lambda  = \left. {\sqrt {\frac{{V''_{{\rm{eff}}} }}{{{\rm{2}}\dot t^2 }}} } \right|_{r_c } \,,
\ee
the approximate QNM frequency is given by
	\be \label{wQNQM}
	\omega _{QNM}  = m\Omega _c  - i\left( {n + 1/2} \right)\left| \lambda  \right| \,.
	\ee 
In the equation above, $m$ is the angular momentum of the perturbation and $n$ is the overtone number \cite{Cardoso:2008bp}.
The corresponding $\dot t$ and $\dot \phi$ in (\ref{OmegaCgen}) and (\ref{lambdaGEN}) are those of the null geodesic, i.e. 
	\be \label{tdot.NULL}
	\dot t = \frac{{g_{t\phi } L + g_{\phi \phi } E}}{\Delta } ~~~{\rm and}~~~ - \dot \phi  = \frac{{g_{tt} L + g_{t\phi } E}}{\Delta }\,,
	\ee 
and the $V_{\rm eff}$ is given in (\ref{Veff.null}).

Not only applied to the general $d$-dimensional Meyrs-Perry black holes, the approach for QNM frequency as given in (\ref{wQNQM}) had been employed to various rotating and charged asymptotically flat black holes in several gravitational set up, such as in Einstein-Maxwell-dilaton \cite{Jai-akson:2017ldo} and modified gravity \cite{Wei:2018aft} theories. Similar to the low energy limit of heterotic string theory described in (\ref{action.het}), the gravitational term appearing in the corresponding actions in \cite{Jai-akson:2017ldo} and \cite{Wei:2018aft} is of first order in the curvature, i.e. it is linear in the Ricci scalar. We then find the recipe described in section IV of \cite{Cardoso:2008bp}, where the authors discussed geodesics in a class of stationary and axial symmetric spacetimes with off-diagonal metric components to obtain the angular frequency and Lyapunov exponent in (\ref{wQNQM}), applies to the Kerr-Sen spacetime considered in this paper.  However, it was reported in \cite{Konoplya:2017wot} that the approximate formula for QNM frequency (\ref{wQNQM}) fails to approach the expected results in the case of black holes in Einstein-Gauss-Bonnet theory. This failure comes from the lacking to obtain exact solutions for black hole horizons, as one needs to expand the metric in terms of some coupling constant $\alpha$. 

\subsection{Generalized BKL recipe}
	
The authors of \cite{Buonanno:2007sv} introduced a prescription, which later known as the BKL approach, to estimate the final spin of an electrically neutral black hole resulting from the merger of two black holes. In this BKL approach, the initial condition for black holes before merging can be spinning or spinless. Some of the assumptions made in constructing BKL prescription are the conservation of mass and angular momentum, before and after collision. In general, BKL method can be applied to any neutral rotating asymptotically flat black holes, for example the Kerr-MOG solution \cite{Wei:2018aft}.

As the electromagnetic interaction is considered, some deviations to the test particle geodesic will result in the final spins and light ring calculations. This is first considered in \cite{Jai-akson:2017ldo}, where the BKL recipe is generalized to the case of rotating and charged black holes. In particular, the objects under consideration in \cite{Jai-akson:2017ldo} are the Kerr-Newman (KN) and Kaluza-Klein (KK) black holes which come from the Einstein-Maxwell-(dilaton) theories respectively. It is found that there are some gaps of the final spins between the Kerr-Newman and Kaluza-Klein black holes, but their general behavior is quite similar.
	
Kerr-Sen black hole is also an exact solution of a rotating and electrically charged collapsing object in theory described by action (\ref{action.het}). The spacetime containing this black hole is also asymptotically flat, stationary, and axial symmetric. A Kerr-Sen black hole is completely identified by its mass, spin, and electric charge, just like the Kerr-Newman counterpart. These three parameters are assumed to be conserved during the merger, which has been also considered in \cite{Jai-akson:2017ldo}. Therefore, the possible generalization of orginal BKL recipe for Kerr-Sen merger case can be done by considering the conservation of black hole charges\cite{Jai-akson:2017ldo}, whereas the original BKL prescription covers the conservations of mass and angular momentum only. However, the conservation of black hole's electric charge does not enter the BKL formula for the final spin directly, but it contributes in obtaining angular momentum and ISCO radius of the test object which then appear in the BKL final spin formula. In other words, the final spin formula according BKL prescription has  the same expression in both neutral and electrically charge black holes mergers.

Here we verify the validity of generalized BKL recipe for the case of Kerr-Sen black holes. This is important since sometime one needs to impose more condition in addition to Newtonian limits, hence the equation describing the radial motion as the results of Newtonian and Coulombic interactions
	\be \label{NewtonianCoulombEQTN}
	\frac{{d^2 r}}{{dt^2 }} + \frac{M}{{r^2 }} = \frac{{qQ}}{{m r^2 }}\,,
	\ee 
can be recovered. For example in KK black hole case, the authors of \cite{Jai-akson:2017ldo} need to consider $Q\ll M$ condition to recover (\ref{NewtonianCoulombEQTN}). Recall that in the last equation, $q$ and $m$ are the electric charge and mass of the probe, and $Q$ is the charged of black hole with mass $M$.
	
Outside a Kerr-Sen black hole, a massive and charged probe described by the Lagrangian (\ref{LagrangianTEST}) obeys the equation of motion
	\be \label{geodesic.eqtn}
	\ddot x^\alpha   + \Gamma _{\kappa \lambda }^\alpha  \dot x^\kappa  \dot x^\lambda   =  - \frac{q}{m }F^{\alpha \beta } \dot x_\beta  \,.
	\ee 
The corresponding Newtonian limit is obtained by imposing the condition ${\dot t} \gg {\dot x}^i$, which yields the eq. (\ref{geodesic.eqtn}) reduces to
	\be 
	\ddot x^\alpha   + \Gamma _{00}^\alpha  \dot t^2  =  - \frac{q}{m }g^{\alpha \beta } F_{\beta 0} \dot t\,.
	\ee 
	Further considerations  $r \gg M$ and $r \gg a$ allow us to write an approximation to the last equation as
	\be \label{radial.eqtn.Newtonian.Coulomb}
	\frac{{d^2 r}}{{dt^2 }} + \frac{M}{{r^2 }} = \frac{{qQ}}{{m r^2 }}\,.
	\ee 
It describes the radial motion of a massive and charged object under the influence of another one which is also electrically charged according to Newtonian gravity and Coulomb interaction.
	
In addition to the conservation of mass and angular momentum, the generalized BKL method uses the assumption that the total charge of black holes before and after merger is conserved, i.e. $Q = Q_1  + Q_2$. Recall that the dynamics of two charged and massive bodies obeying Newton and Coulomb laws can be written as
	\be \label{eqtnM12Q12}
	\left( {\frac{{M_1 M_2 }}{{M_1  + M_2 }}} \right)\frac{{d^2 r}}{{dt^2 }} + \frac{{M_1 M_2 }}{{r^2 }} = \frac{{Q_1 Q_2 }}{{r^2 }}\,.
	\ee 
To recover the last equation from eq. (\ref{radial.eqtn.Newtonian.Coulomb}), in addition to the reduced mass 
\be \label{probe.mass}
m = \frac{{M_1 M_2 }}{{M_1  + M_2 }}
\ee 
we need also to set the probe charge 
	\be \label{probe.charge}
	q = \frac{{Q_1 Q_2 }}{{Q_1  + Q_2 }}\,.
	\ee 
Then one can find that (\ref{eqtnM12Q12}) is just (\ref{radial.eqtn.Newtonian.Coulomb}) with $m$ and $q$ as given in (\ref{probe.mass}) and (\ref{probe.charge}), respectively. To simplify our plots in the coming sections, let us assign $\xi = Q_2 / Q_1$ as the ratio of two black hole charges. As the result, the charge to mass ratio of probe in the equal masses consideration can have the form
	\be 
	e = \frac{{4\xi Q^* }}{{\left( {1 + \xi } \right)^2 }}\,.
	\ee 
In the equation above we have used $Q^*$ to represent the ratio of black hole charge $Q$ to its mass $M$. In the next sections, we will also employ this ``starred'' notation to expressed the ratio of some other quantities of the black hole to its mass.

In the generalized BKL approach, we still employ the original form of angular momentum conservation of BKL method
\be\label{BKLeq1}
M A_f  = L_{{\rm{test}}} \left( {r_{ISCO} ,A_f } \right) + M_1 A_1  + M_2 A_2 \,,
\ee 
to predict the final spins of rotating and charged black holes. Above, $A_1$ and $A_2$ are spin parameters of the two initial black holes, and $A_f$ is the final spin of black hole resulting from the merger. The different feature now for the case of charged black holes is that the probe's angular momentum $L_{{\rm{test}}} \left( {r_{ISCO} ,A_f } \right)$ gets contribution from the Coulomb interaction between the probe and the black hole. This probe has mass $m$ (\ref{probe.mass}) and charge $q$ (\ref{probe.charge}) and orbiting near the black hole. Note that the test particle angular momentum is evaluated at $r_{ISCO}$, and the final spin $A_f$ normally can be found by solving eq.  (\ref{BKLeq1}). The last formula can be rewritten in a more elegant form by defining $\chi _1  = A_1 M_1^{ - 1} $, $\chi _2  = A_2 M_2^{ - 1} $, and $\nu  = m M^{-1}$, which leads to \cite{Buonanno:2007sv}
\be \label{BKLeq2}
M A_f  = \nu L_{{\rm{test}}} \left( {r_{ISCO} ,A_f } \right) + \frac{{M^2 }}{4}\left( {\chi _1 \left( {1 + \sqrt {1 - 4\nu } } \right)^2  + \chi _2 \left( {1 - \sqrt {1 - 4\nu } } \right)^2 } \right)\,.
\ee 
The final spin of black hole after merger is obtained by solving the last equation.

A special case that one can consider is black holes with equal spin parameters. In such consideration we have $\chi_1 = \chi_2 = \chi$, and the reading of eq. (\ref{BKLeq2}) becomes
\be \label{AfEqualspin}
A_f  = L^*\left( {r_{ISCO} ,A_f } \right)\nu  + M\left( {1 - 2\nu } \right)\chi  \,.
\ee 
Again, we have used our convention starred convention where $L^* = L/M$. Furthermore, one can also think of a simpler case where the two black holes are initially nonspinning with equal initial mass. In this consideration, eq. (\ref{AfEqualspin}) reduces to 
\be\label{AfEqualmassspin}
A_f = \tfrac{1}{4} L^*\left( {r_{ISCO} ,A_f } \right)\,.
\ee 
In employing the generalized BKL recipe, note that the conservation of electric charge does not directly incorporated in BKL formula (\ref{BKLeq1}).

\section{Merger estimates in pure geodesics}\label{s4}
	
\subsection{Equatorial orbits}
	
In pure geodesic we consider binary black holes system which finally collide consists of a neutral and an electrically charge black holes. Therefore, there is no Coulomb interaction between the two black holes. However, one of the black hole is still charged, then some deviations to the case of two neutral Kerr black holes merger are expected. The system of two charged black holes will be our discussion in the next section. Since we are considering the motion of an object which lies on the equatorial plane\footnote{In \cite{Siahaan:2015ljs} we showed that this motion exists in Kerr-Sen background.}, the associated metric tensor components are
	\be 
	g_{tt}  = {-1 + \frac{{2M}}{{r + 2b}}}
	~,~
	g_{rr}  = \frac{r^2 + 2br}{{\Delta  }}
	~,~
	g_{t\phi }  =  - \frac{{2Ma}}{{r + 2b}}
	~,~
	g_{\phi \phi }  = r^2 + 2br + a^2  + \frac{{2Ma^2 }}{{r + 2b}}\,.
	\ee 
Here the identity $\Delta  = g_{t\phi }^2  - g_{tt} g_{\phi \phi } $ holds, and is vital in our formula derivations. The corresponding conserved quantities in pure geodesic set up are those in eqs. (\ref{Egen}) and (\ref{Lgen}) with $e=0$,
	\be 
	- E = g_{tt} \dot t  + g_{t\phi } \dot \phi ~~{\rm and}~~ L = g_{t\phi } \dot t  + g_{\phi \phi } \dot \phi \,
	\ee
which leads to the effective potential
	\be 
	V_{{\rm{eff}}}  = \frac{{g_{tt} L^2  + 2g_{t\phi } LE + g_{\phi \phi } E^2  - \Delta }}{{g_{rr} \Delta }}\,.
	\ee 
The last formula is just eq. (\ref{VeffTIMELIKE}) for $e=0$.
	
Solving the two equations $ V_{{\rm{eff}}}  = 0$ and $V'_{{\rm{eff}}}  = 0$ simultaeously, one can obtain the solutions for test particle energy and angular momentum
	\be \label{LpmPUREtimlike}
	L_ \pm   = \frac{{ \pm \sqrt M \left( {a^2  + r\left( {r + 2b} \right) \mp 2\sqrt {M\left( {r + b} \right)} a} \right)}}{{\left( {\left( {r + 2b} \right)\left( {\left( {r + b} \right)^2  - 3M\left( {r + b} \right) + \left( {r + M + b} \right)b \pm 2a\sqrt {M\left( {r + b} \right)} } \right)} \right)^{1/2} }}\,,
	\ee 
	and
	\be \label{EpmPUREtimlike}
	E_ \pm   = \frac{{\left( {r + b} \right)^{3/2}  + \left( {b - 2M} \right)\sqrt {r + b}  \pm a\sqrt M }}{{\left( {\left( {r + 2b} \right)\left( {\left( {r + b} \right)^2  - \left( {r + b} \right)\left( {3M - b} \right) + Mb \pm 2a\sqrt {M\left( {r + b} \right)} } \right)} \right)^{1/2} }}\,,
	\ee
where the subscripts $+$ and $-$ stand for the prograde or direct and retrograde motions. Furthermore, the ISCO radius can be found by solving $V_{\rm eff}'' =0$, i.e.
	\be \label{dVVpure}
	u^6  + \left( {3b - 6M} \right)u^4  \pm 8a\sqrt M u^3  + 3\left( {b^2  - a^2 } \right)u^2  - 2Mb^2  - a^2 b = 0
	\ee 
where $u^2 = r+b$. Setting $b \to 0$, which is used to transform Kerr-Sen solution to Kerr, yields eq. (\ref{dVVpure}) reduces to the equation for marginally stable orbit in Kerr background \cite{Bardeen:1972fi}
	\be 
	r\left( {r - 6M} \right) \pm 8a\sqrt {rM}  - 3a^2  = 0\,,
	\ee 
Furthermore the ISCO radius of \sch black hole $r=6M$ \cite{Bardeen:1972fi}, is the solution of equation (\ref{dVVpure}) evaluated at $a \to 0$ and $b \to 0$.

\subsection{Light ring}

Here we will obtain the Lyapunov exponent $\lambda$ and angular velocity $\Omega_{\rm c}$ related to null geodesic in Kerr-Sen geometry which are needed to approximate QNM frequencies using eq. (\ref{wQNQM}). In the null geodesic, equation $V_{\rm eff}=0$ gives us the ratio between the test particle energy to its angular momentum,
	\be 
	\frac{{E_ \pm  }}{L} = \frac{{2Ma \pm \left( {r + 2b} \right)\sqrt \Delta  }}{{r\left( {r + 2b} \right)^2  + a^2 \left( {r + 2M + 2b} \right)}} \,.
	\ee 
Inserting this ratio to the expression of $\dot t$ in (\ref{tdot.NULL}) leads to $\dot \phi  =  \pm \Delta ^{ - 1/2} L$, where the upper and lower signs refer to the solution for prograde and retrograde respectively. Evaluating the general Lyapunov formula (\ref{lambdaGEN}) in this null geodesic gives
	\be \label{Lyapunov}
	\lambda = \frac{{{a^*}{{\left( {{a^*}^2\left( {3{b^*} - 3 - {b^*}^2} \right) + \left( {3 - 2{b^*}} \right){{\left( {1 - {b^*}} \right)}^2}} \right)}^{1/2}}}}{{2M\left( {2 - 2{b^*} + {a^*}^2} \right)\left( {1 - {b^*}} \right)}}\,,
	\ee 
and the angular velocity at circular radius $r_{\rm c}$ is
	\be \label{velocity.ang}
	\Omega _c  = \frac{a^*}{M\left({{a^*} ^2 -2 b^* +2}\right)}\,.
	\ee 
The last two equations have been expressed in terms of the ratios of black hole's rotational parameter $a$ and electric charge $b$ to the black hole's mass $M$, i.e. $a^* = a/M$ and $b^* = b/M$. Note that the two formulas $\lambda$ and $\Omega_c$ above are also valid for the charged geodesic consideration later on. In evaluating (\ref{Lyapunov}) and (\ref{velocity.ang}) numerically, the value of $a^*$ is the final spin $A_f ^*$ obtained by solving eq. (\ref{BKLeq2}). Here we can understand how the charge of black holes or Coulomb interaction gives contribution to the approximate QNMs frequency using the light ring. Some numerical results for $\lambda$ dan $\Omega_{\rm c}$ are given in section \ref{s5}.

\subsection{Final spins}

The extremal limit of a Kerr-Sen black hole is $a = M-b$, i.e. a naked singularity is produced when $a > M-b$. Hence, for the merger of two Kerr-Sen black holes with final mass $M = M_1 +M_2$ and final charge $Q = Q_1 +Q_2$, the relation
\be\label{extreme.bound}
M - \frac{{Q^2 }}{{2M}} \ge \left| {A_f } \right|
\ee
must be fulfilled to avoid the production of a naked singularity. Above, $A_f$ is rotational parameter or spin of the final black hole. It is obvious from the last inequality that the maximal spin $A_f$ decreases as the charge parameter $b = Q^2/2M$ increases\footnote{The same conclusion is also drawn in Kerr-Newman case \cite{Jai-akson:2017ldo} whose black hole condition is $M^2-Q^2 \ge \left| {A_f } \right|$.}. In addition to our concern on the possibility of naked singularity formation after black hole merging, we also need to put a constraint for the charge to mass ratio $e$ that keeps the non-spinning test particle not becoming a naked singularity. The corresponding constraint is $2m^2 \ge q^2$ which gives $ \left| e \right| \le \sqrt{2} $. 

\begin{figure}
	\centering
	\includegraphics[scale=0.5]{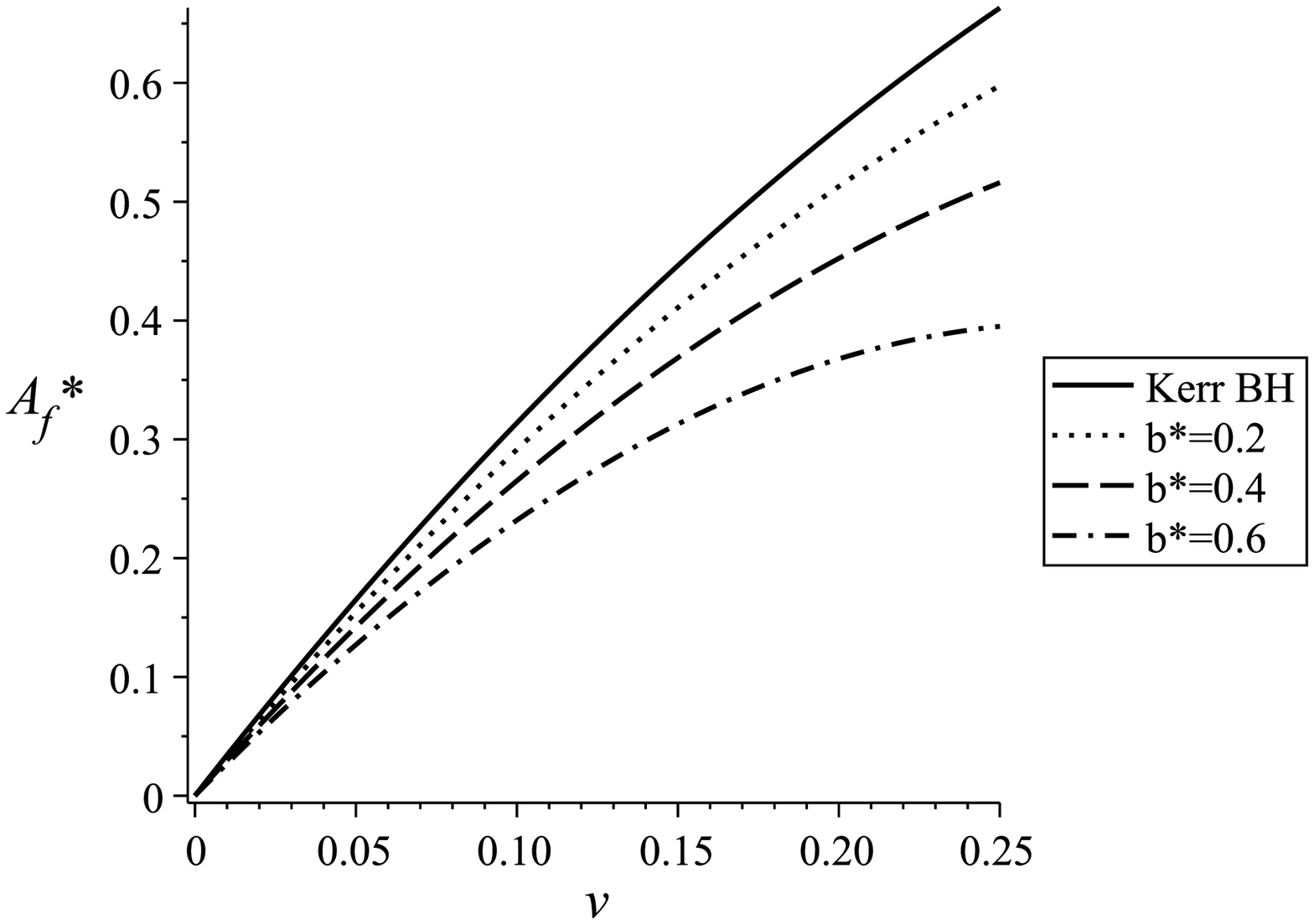}
	\caption{$\chi = 0$.}\label{fig.afMtoVforX0}
\end{figure}

\begin{figure}
	\centering
	\includegraphics[scale=0.5]{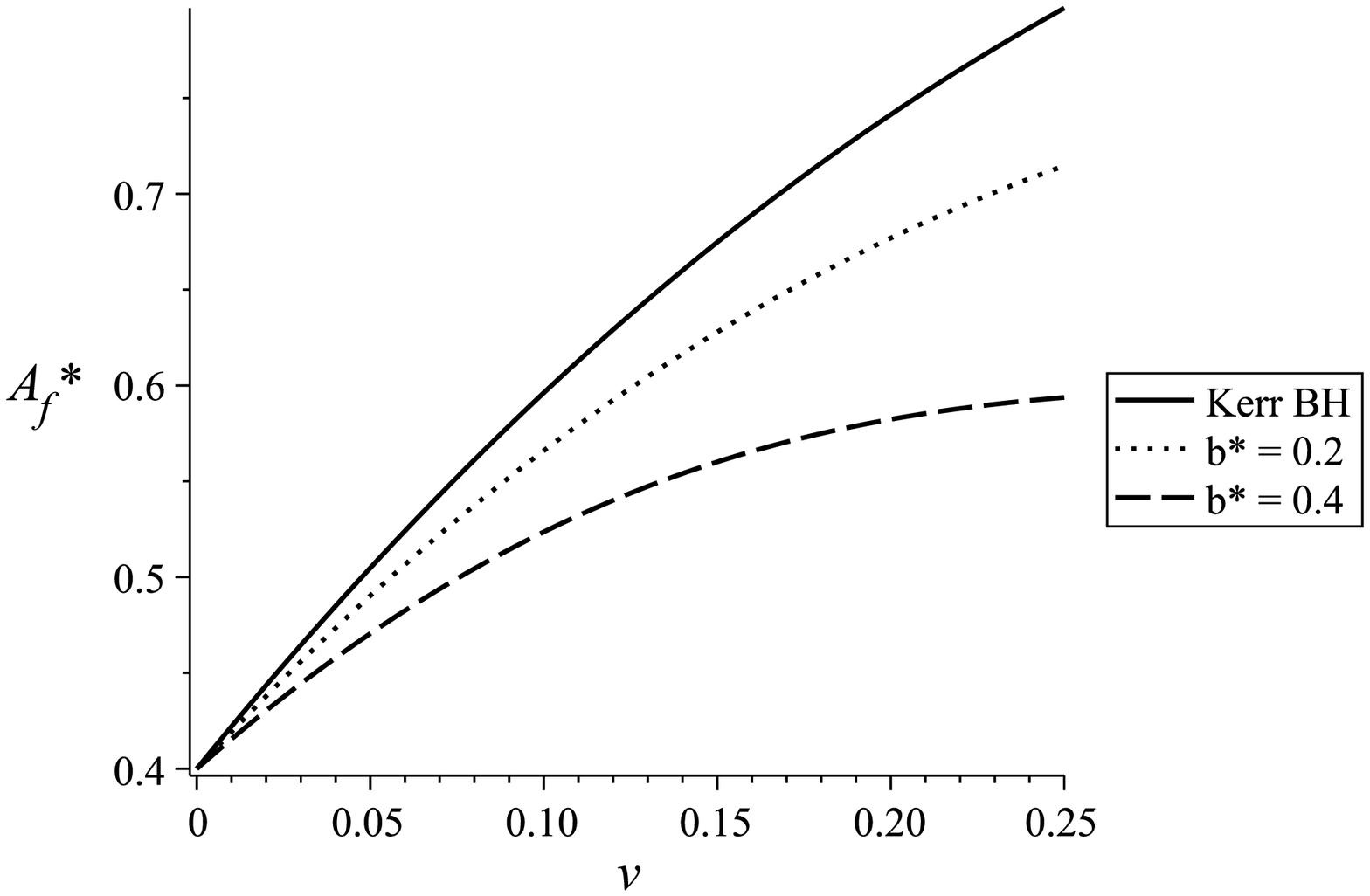}
	\caption{$\chi = 0.3$.}\label{fig.afMtoVforX03}
\end{figure}

To give illustrations for black hole mergers in pure geodesic consideration, here we provide some plots representing the merger of two black holes with equal initial spins. The plots are presented as the final spins against $\nu$, which tell us how the final spin varies as the initial masses ratio between the two black holes changes\footnote{Let $\zeta$ is the ratio of initial black hole masses, i.e. $M_2 = \zeta M_1$, then one can show $\nu  = M_1 M_2 M^{ - 2}  = \zeta \left( {1 + \zeta } \right)^{ - 2} $.}. The cases of initial spins $\chi =0$ and $\chi = 0.3$ are showed in figs. \ref{fig.afMtoVforX0} and \ref{fig.afMtoVforX03}, respectively. From these plots, we learn that the final spins grow as the ratio between masses of black holes increases. Note that the maximum of real $\nu$ is $\nu = \tfrac{1}{4}$ for the equal initial masses of black holes. Interestingly in both plots \ref{fig.afMtoVforX0} and \ref{fig.afMtoVforX03}, we observe that two black holes system with the same mass configuration $\nu$ but larger electric charge end up with smaller spin after merger. In other words, the maximum final spin is obtained for neutral black holes system, i.e. Kerr black holes. The similar findings also appear in the study of Kerr-Newman black holes \cite{Jai-akson:2017ldo}. 
	
\begin{figure}
	\centering
	\includegraphics[scale=0.5]{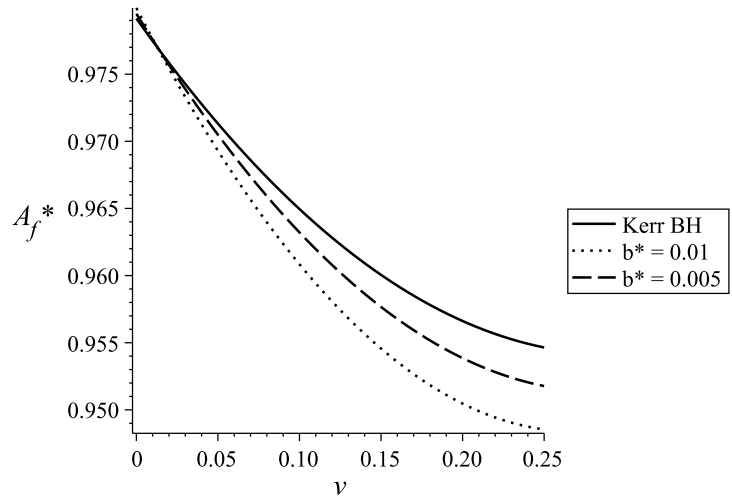}
	\caption{$\chi = 0.98$.}\label{fig.afMtoVforX098}
\end{figure}

An interesting behavior comes out in fig. \ref{fig.afMtoVforX098}. For two black holes with a large initial rotational parameter, i.e. $\chi  = 0.98$, the final spin after merger decreases as $\nu$ increases. Nevertheless, the bound $A_f ^* +b^* \le 1$ which is just a rewriting of (\ref{extreme.bound}) is still satisfied in this extreme consideration. Similar to the case of $\chi = 0$ and $\chi = 0.3$, in fig. \ref{fig.afMtoVforX098} we observe that the presence of electric charge lowers the final spin.

\section{Merger estimates for charged and rotating black holes}\label{s5}
	
\subsection{Final spins}
	
\begin{figure}
		\centering
		\includegraphics[scale=0.5]{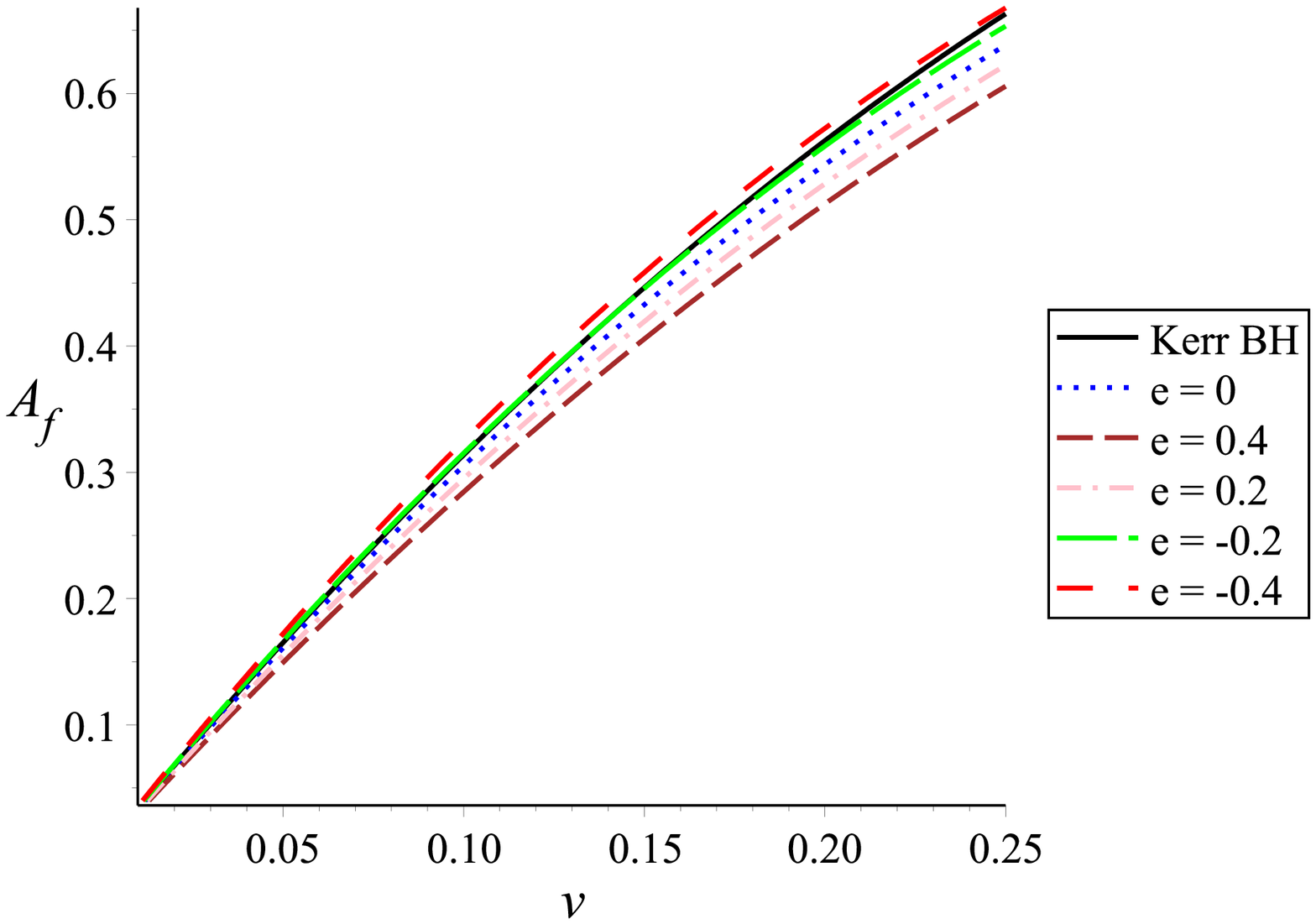}
		\caption{The case of $\chi = 0$ and $Q^* = 0.4$.}\label{fig.AfMtovChargedX0}
	\end{figure}

In this section we deal with binary Kerr-Sen black holes. Therefore we expect some corrections to the pure geodesics coming from the electromagnetic interaction between the black holes. The corresponding effective potential in this case is that of eqs. (\ref{VeffTIMELIKE}). Similar to the pure geodesic consideration, the three equations $V_{\rm eff}=0$, $V'_{\rm eff}=0$, and $V''_{\rm eff}=0$ constraint the energy $E$, angular momentum $L$, and ISCO radius $r_{\rm ISCO}$ of a probe. Expressing the exact form of these three quantitites for Kerr-Sen spacetime is not so easy, and yet it is not straightforward to extract some qualitative information from them. For these reasons, again we turn to numerical plots in showing how the final spins of black holes vary to the change of $\nu$. In making these plots, we examine the equal initial spins only, i.e. $\chi = 0$ in fig. \ref{fig.AfMtovChargedX0} and $\chi = 0.4$ in fig. \ref{fig.AfMtovChargedX0}. Both plots are the case of Kerr-Sen black holes with $Q^* = 0.4$ or equivalently $b^* = 0.08$.

	\begin{figure}
		\centering
		\includegraphics[scale=0.5]{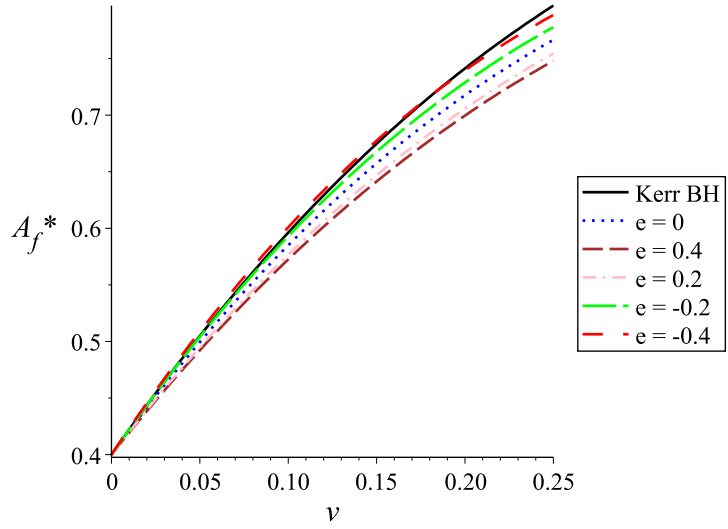}
		\caption{The case of $\chi = 0.4$ and $Q^* = 0.4$.}\label{fig.AfMtovChargedX04}
	\end{figure}
	
\begin{figure}
	\centering
	\includegraphics[scale=0.5]{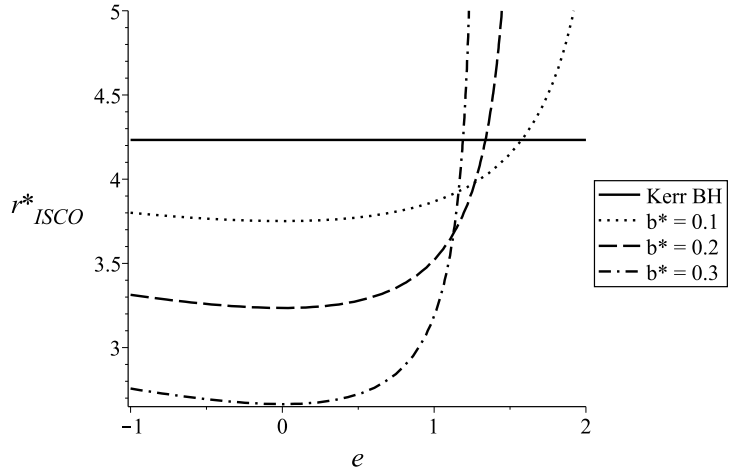}
	\caption{ISCO radius per unit mass varies with respect to charge  to mass of the probe, $e$. The plot is evaluated at $A^* = 0.5$.}\label{fig.riscotoe}
\end{figure}

\begin{figure}
	\centering
	\includegraphics[scale=0.5]{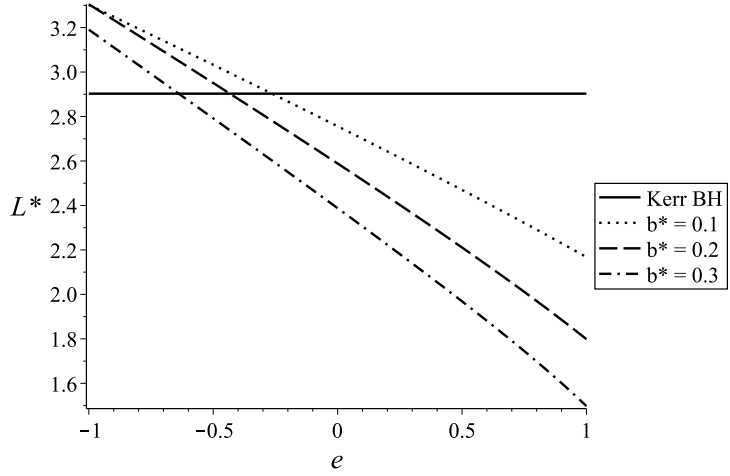}
	\caption{Probe angular momentum per unit mass $L^*$ vs. $e$ evaluated at $r_{\rm ISCO}$ and for $A^* = 0.5$.}\label{fig.LMtoe}
\end{figure}

From fig. \ref{fig.AfMtovChargedX0} we learn that final spin tends to decrease as the charge to mass ratio of the probe $e$ increases. Even when $e = -0.4$, the final spin takes values bigger that of Kerr black holes. The same conclusion can also be drawn from the equal initial spin $\chi = 0.4$ case as depicted in fig. \ref{fig.AfMtovChargedX04}. For the case of nonspinning initial case, the maximum final spin for equal initial masses for black holes occurs for $e = -0.4$. On the other hand, when for the initial spin $\chi = 0.4$, the maximum final spin is that of Kerr black holes if the initial masses are equal. I addition to the plots of final spins against $\nu$ in figs. \ref{fig.AfMtovChargedX0} and \ref{fig.AfMtovChargedX04}, we also present graphics in figures \ref{fig.riscotoe} and \ref{fig.LMtoe} describing how the ISCO radius and probe angular momentum vary with respect to charge to mass ratio of probe $e$. We find that the behaviors of $r^*_{\rm ISCO}$ and $L^*$ for a probe outside a Kerr-Sen black hole are similar to that of Kerr-Newman \cite{Jai-akson:2017ldo}.
	
\subsection{Light ring}

\begin{figure}
	\centering
	\includegraphics[scale=0.5]{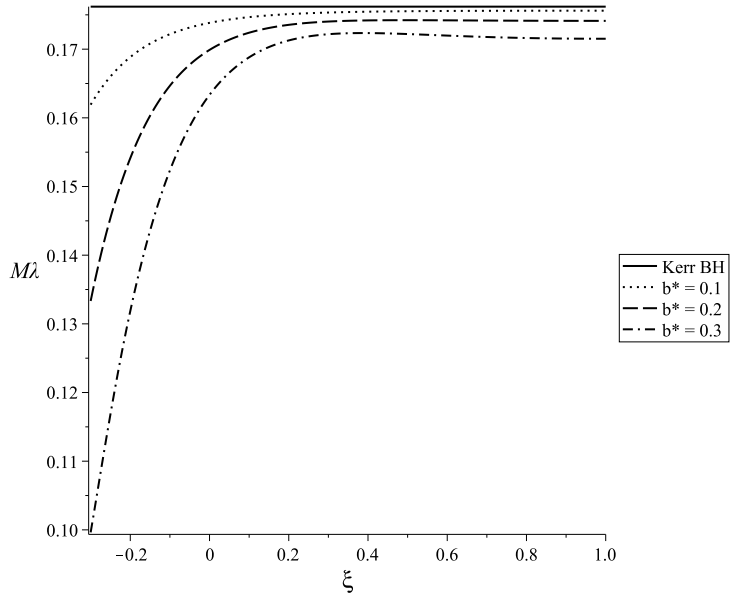}
	\caption{$M\lambda$ of light ring vs. ratio of initial black holes charges. Evaluated for initial spin $\chi =0$.}\label{fig.LambdaMtoQ2Q1}
\end{figure}

\begin{figure}
	\centering
	\includegraphics[scale=0.5]{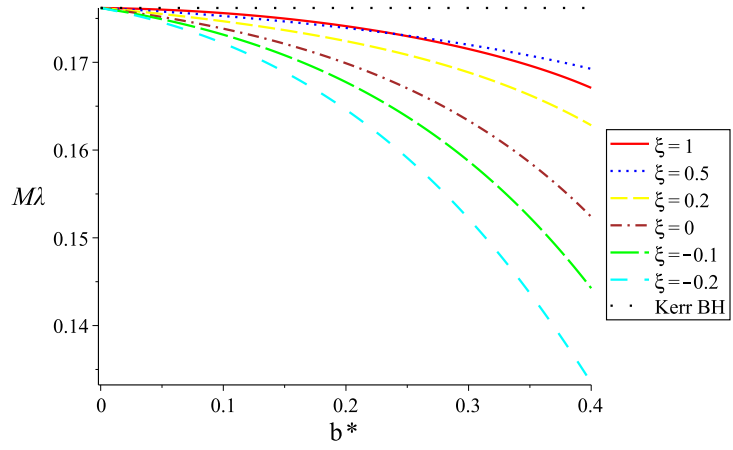}
	\caption{$M\lambda$ of light ring vs. final black hole's charge. Evaluated for initial spin $\chi =0$.}\label{fig.LambdaMtoQoM}
\end{figure}

\begin{figure}
	\centering
	\includegraphics[scale=0.5]{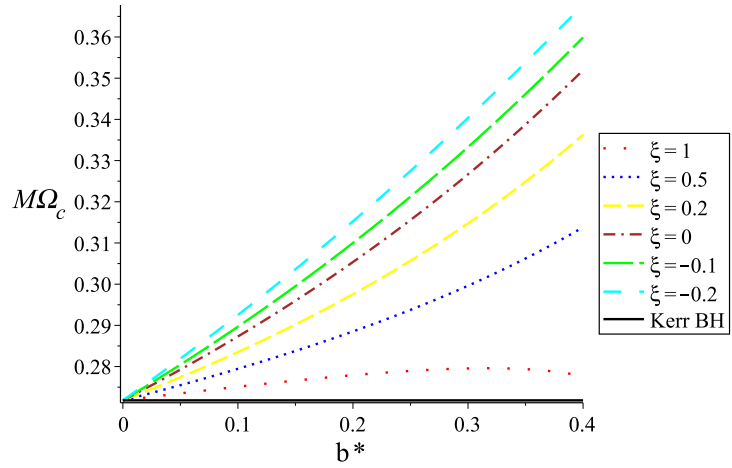}
	\caption{$M\Omega_c$ of light ring vs. final black hole's charge. Evaluated for initial spin $\chi =0$.}\label{fig.MWctoQoM}
\end{figure}

\begin{figure}
	\centering
	\includegraphics[scale=0.5]{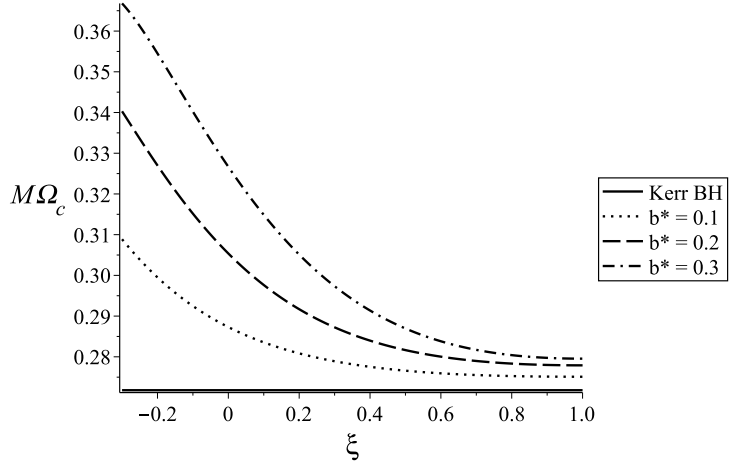}
	\caption{$M\Omega_c$ of light ring vs. the ratio of initial black hole's charge. Evaluated for initial spin $\chi =0$.}\label{fig.MWctoQ2oQ1}
\end{figure}

Now let us provide numerical results for Lyapunov exponent (\ref{Lyapunov}) and angular velocity (\ref{velocity.ang}) evaluated at the final spin of black holes. The final spin of black holes is dictated by the formula (\ref{AfEqualmassspin}) since we consider the initial nonspinning case only. In general, the behavior of Lyapunov exponent for null object near the Kerr-Sen black holes is similar to that of Kerr-Newman. We observe in fig. \ref{fig.LambdaMtoQ2Q1} that $\lambda$ grows as black holes charge ratio $\xi$ raises. In fig. \ref{fig.LambdaMtoQoM}, we also notice that $\lambda$ decreases for the larger final charge of black holes. However, we find several discrepancies which are the followings. In Kerr-Newman case \cite{Jai-akson:2017ldo} the plots of $M\lambda$ vs. $\xi$ can intersect at some values of $\xi$, which is not resembled in fig. \ref{fig.LambdaMtoQ2Q1}. Another difference is for the equal initial charge of Kerr-Newman black holes, i.e. $\xi = 1$, $M\lambda$ can be bigger than that of Kerr black holes\footnote{This behavior also appears slightly for $\xi = 0.5$.} in some range of $\xi$. In fig. \ref{fig.LambdaMtoQoM}, we find that the maximum $M\lambda$ is that of Kerr black holes. For angular velocity plots in figs. \ref{fig.MWctoQoM} and \ref{fig.MWctoQ2oQ1}, we can see resemblances to the Kerr-Newman case. The angular velocity $\Omega_{\rm c}$ grows as the final black hole charge increases, and it decreases as the black holes initial charge ratio goes to unity.

\begin{figure}
	\centering
	\includegraphics[scale=0.5]{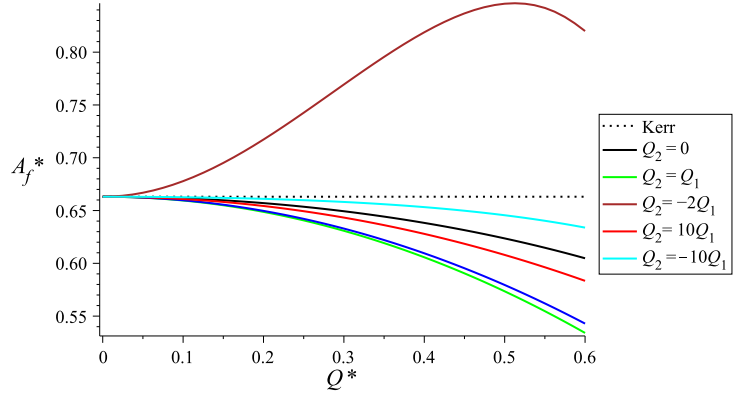}
	\caption{Final spin of black hole vs. final black hole's charge. Evaluated for initial spin $\chi =0$.}\label{fig.AftoK-KS}
\end{figure}

\begin{figure}
	\centering
	\includegraphics[scale=0.5]{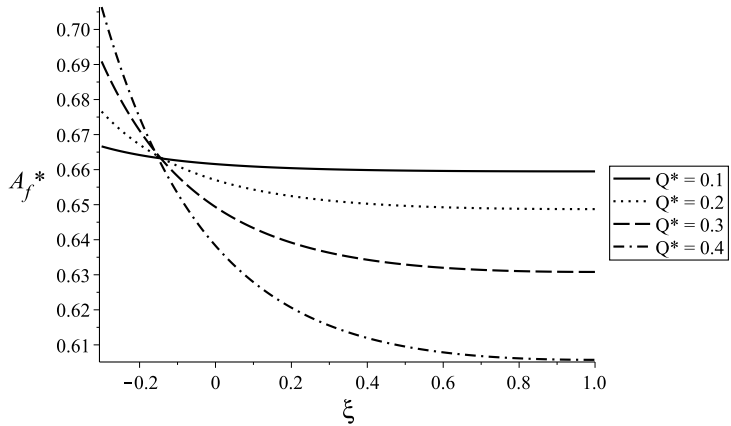}
	\caption{Final spin of black hole vs. the ratio of initial black holes charge. Evaluated for initial spin $\chi =0$.}\label{fig.AftoX}
\end{figure}
	
In fig. \ref{fig.AftoK-KS} we provide some plots for black hole's final spin varies with respect to the final charge, while in fig. \ref{fig.AftoX} the final spin is plotted against the ratio of initial black hole's charge. In making these plots, we consider the initial nonspinning consideration. Fig. \ref{fig.AftoK-KS} tells us that the final spin decreases as charge raises, even when $Q_2 = - 10 Q_1$. However, a quite peculiar behavior is noticed for the case $Q_2 = -2Q_1$, i.e. the two black holes are oppositely charged but the magnitude is comparable between each other. The final spin initially grows as the total charge increases and falls after reaching a peak point. On the other hand, from fig. \ref{fig.AftoX} we learn that the final spin of black hole decreases as the charge ratio between the two black hole gets larger. 
	
\section{Kerr-Newman and Kerr-Sen cases}\label{s6}
	
In this section we provide some plots showing the final spins  resulting from the black hole merger, for the case of Kerr-Sen and Kerr-Newman. As we have mentioned, the two black holes are quite similar in some aspects. Using the standard textbook formula \cite{Wald:1984rg} to get the mass, angular momentum, and electric charge, one can obtain $M$, $J=Ma$, and $Q$, respectively, for the metric and vector solutions in eqs. (\ref{KerrSenmetricEinsteinFrame}), (\ref{KerrSenVectors}), (\ref{metricKN}), and (\ref{gaugeKN}). However keep in mind that Kerr-Sen solution comes from the low energy limit of heterotic string theory, while Kerr-Newman solution belongs to the Einstein-Maxwell theory. Since the low energy limit of heterotic string theory is an alternative low energy gravity description to the Einstein-Maxwell framework, one may wonder how differ the two in modeling the merger of rotating and charged black holes using the generalized BKL formalism. We can consider a simple case where the initial spins are zero, and the two merging black holes have equal initial masses, and also equal initial electric charge. In this consideration, we have $\chi =0$, $e= Q^*$, and $\nu = 0.25$, and we can rely on the eq. (\ref{AfEqualspin}).
	
\begin{figure}
	\centering
	\includegraphics[scale=0.5]{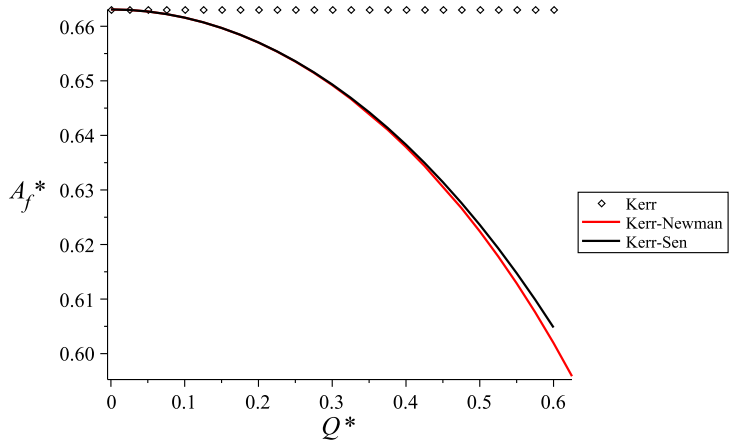}
	\caption{The case of $\chi=0$.}\label{fig.AftoK-Q2-0}
\end{figure}

\begin{figure}
	\centering
	\includegraphics[scale=0.5]{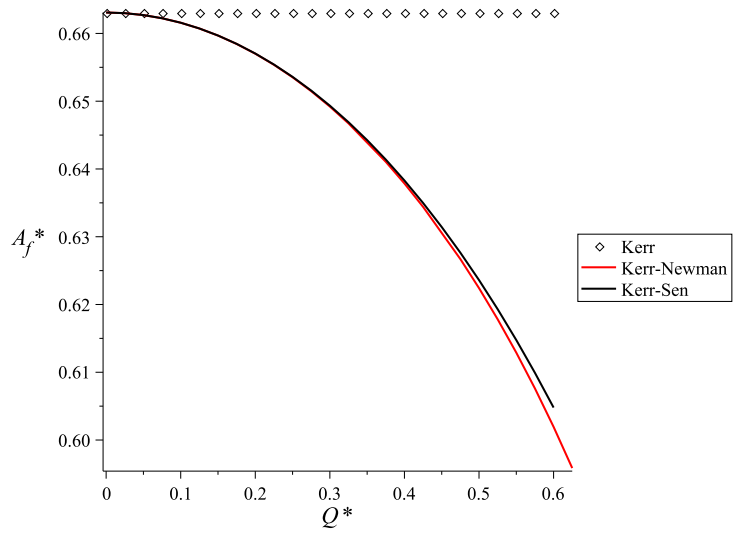}
	\caption{The case of $\chi=1$.}\label{fig.AftoK-Q2-Q1}
\end{figure}

\begin{figure}
	\centering
	\includegraphics[scale=0.5]{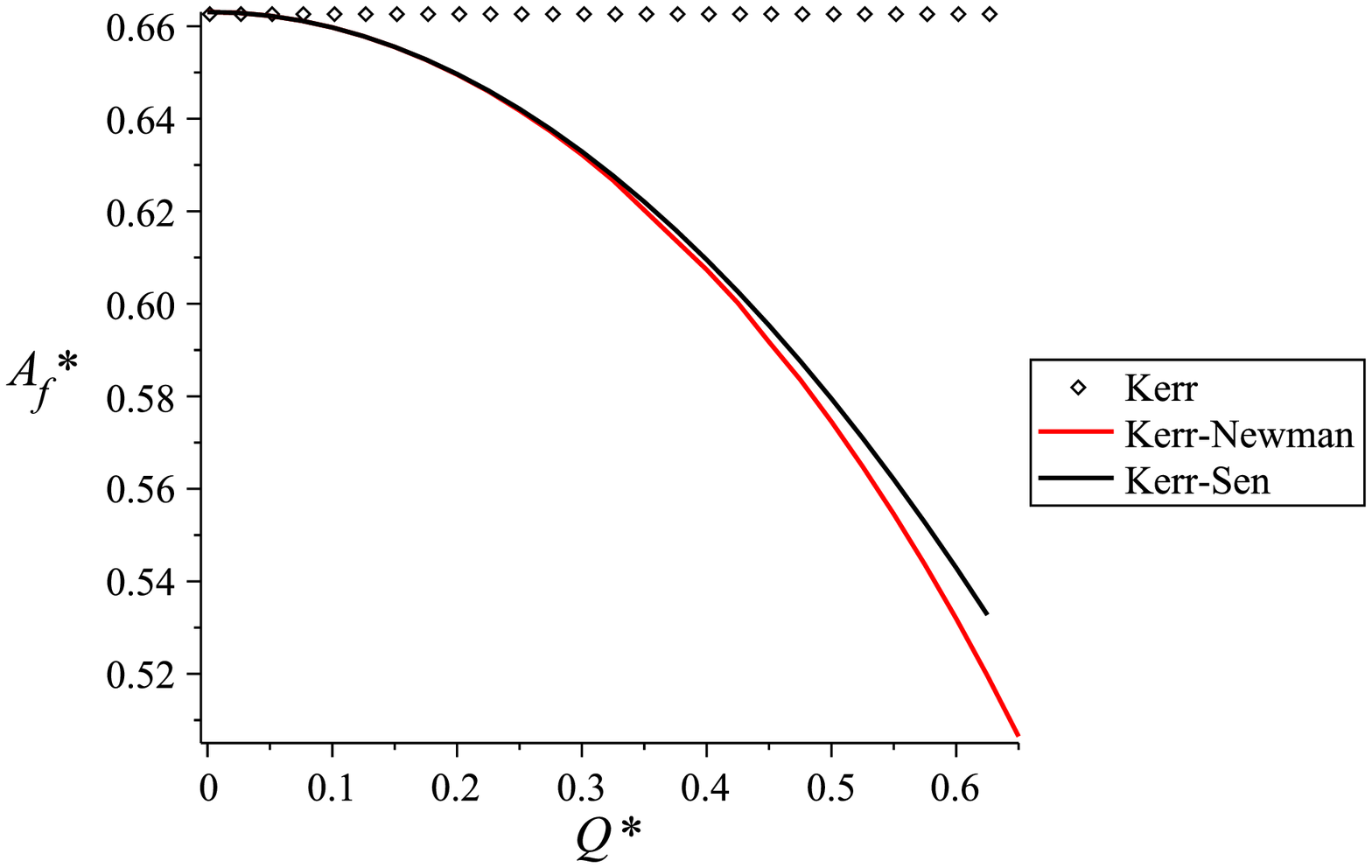}
	\caption{The case of $\chi=2$.}\label{fig.AftoK-Q2-2Q1}
\end{figure}

\begin{figure}
	\centering
	\includegraphics[scale=0.5]{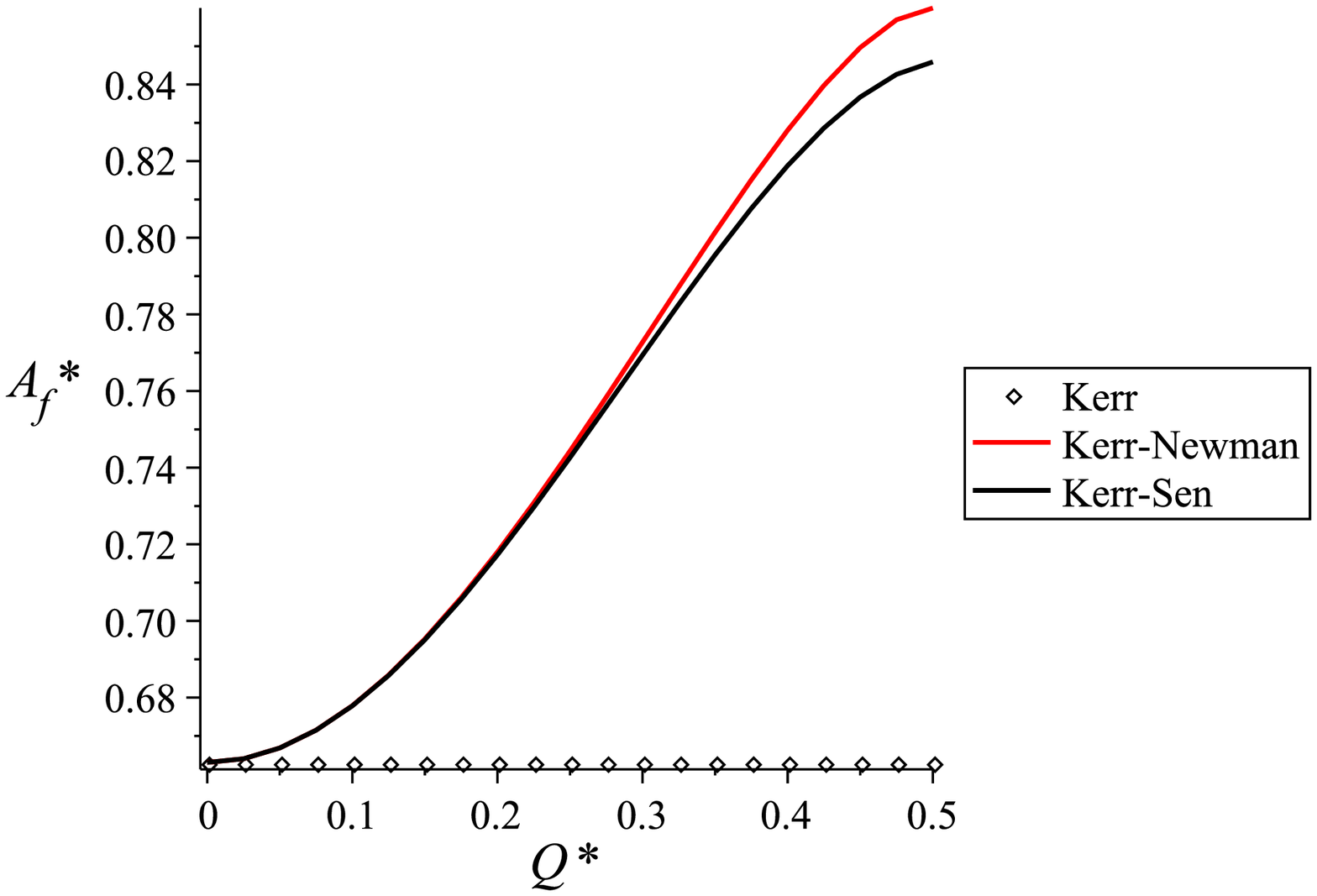}
	\caption{The case of $\chi=-2$.}\label{fig.AftoK-Q2-min2Q1}
\end{figure}

The final spin plots are given in figs. \ref{fig.AftoK-Q2-0}, \ref{fig.AftoK-Q2-Q1}, \ref{fig.AftoK-Q2-2Q1}, and \ref{fig.AftoK-Q2-min2Q1}, which represent the case several initial spins $\chi = 0$, $\chi = 1$, $\chi = 2$, and $\chi = -2$, respectively. It is observed that in all cases considered in these four plots, the general behavior of final spins after merger depending on the final charge of black hole is similar between Kerr-Newman and Kerr-Sen black holes. Especially in the regime of smaller total charge, the final spins per unit mass of the two black holes are quite overlapping. This can be understood since the neutral limit of Kerr-Sen and Kerr-Newman black holes are the same, i.e. Kerr solution. As the total charge per unit mass $Q^*$ raises, a gap between the final spin of the two black hole starts to increase.

\section{Conclusions}\label{s7}
	
In this paper we investigate the coallesence of two Kerr-Sen black holes. We use an approach presented in \cite{Jai-akson:2017ldo} where the authors studied the mergers of Kerr-Newman and Kaluza-Klein black holes. The method can be considered as a generalization to the BKL formalism \cite{Buonanno:2007sv}, where the authors of \cite{Jai-akson:2017ldo} added the electromagnetic interaction into consideration. As suggested in \cite{Jai-akson:2017ldo}, their method should apply to any rotating and charged black holes provided that the exact expression of the spacetime metric and gauge field are known. 

In general, the estimates behavior of black holes merger in Kerr-Sen case is quite similar to that of Kerr-Newman, as it is expected. For example in section \ref{s6}, we notice that final spins are lowered as the total black hole charge grows for initial spins $\chi = 0,1,2$, and increases for initial spin $\chi = -2$. This conclusion applies to Kerr-Sen and Kerr-Newman black holes. However, we notice that there also exist some slight distinctions as pointed in sections \ref{s5} and \ref{s6}. For example there is no intersection for plots in fig. \ref{fig.LambdaMtoQ2Q1}, and the Lyapunov exponent $\lambda$ multiplied by black hole mass $M$ of Kerr-Sen case is below that of Kerr black hole for any final back hole charge to mass ratio $b^*$.

In this paper, we limit our studies to the equatorial case and neglecting the spin-spin interaction between the merging black holes. However, in a more generic case, two black holes with arbitrary spin directions could lead to the precession of orbital plane and the spins contribute to the final angular momentum. To include the spin-spin interaction in rotating and charged black holes binary merger, one can employ the prescription proposed in \cite{Buonanno:2007sv} to generalize the BKL formula (\ref{BKLeq1}) for the case of constant angle between total spin and angular momentum. As electromagnetic interaction does not appear directly in the formula (\ref{BKLeq1}) to estimate the final spin, the same would apply if one considers the spin-spin interaction for rotating charged black holes merger using the prescription in section III.C of \cite{Buonanno:2007sv}. The electromagnetic interaction comes into play in the part when we ought to obtain the ISCO radius or angular momentum of the test body by solving the set of equations $V_{\rm eff}=0$, $V'_{\rm eff}=0$, and $V''_{\rm eff}=0$. To the best of our knowledge, the generalized BKL prescription with spin-spin interaction consideration to estimate the final spin of merged spinning and charged black holes has never been worked out in literature. We left this project as our future work. 

There exist some other exact rotating and charged black hole solutions that belong to some gravitational theory beyond Einsteins available in literature, for example in \cite{Ortin:2015hya}. These solutions are worth to be investigated, hence the results can be confronted to the forthcoming data on black hole properties \cite{Barack:2018yly} with more precisions. Related to the electromagnetic interaction which is considered in the generalized BKL method \cite{Jai-akson:2017ldo}, it is also interesting to see how significant the contribution of external magnetic fields \cite{Siahaan:2016zjw,Siahaan:2015xia} to the final spin of black holes and the QNMs frequencies.
	
\section*{Acknowledgement}
	
This work is supported by LPPM-UNPAR under the contract no. PL72018028. I thank Reinard Primulando, Puttarak Jai-akson, and Paulus Tjiang for useful discussions.

\appendix
\section{Kerr-Newman solution}
	
The Einstein-Maxwell action reads
	\be \label{action.EM}
	S = \int {d^4 x\sqrt { - g} \left( {R - \frac{1}{4}g^{\alpha \mu } g^{\beta \nu } F_{\alpha \beta } F_{\mu \nu } } \right)} \,.
	\ee 
The corresponding line element is
	\[
	ds^2  =  - \left( {1 - \frac{{2Mr - Q^2 }}{{\bar \rho ^2 }}} \right){\rm{d}}t^2  - \frac{{2\left( {2Mr - Q^2 } \right){\rm{d}}t{\rm{d}}\phi }}{{\bar \rho ^2 }} + \bar \rho ^2 \left( {\frac{{{\rm{d}}r^2 }}{{\bar \Delta }} + \frac{{{\rm{d}}x^2 }}{{\left( {1 - x^2 } \right)}}} \right)
	\]
	\be \label{metricKN}
	+ \frac{{\left( {1 - x^2 } \right)}}{{\bar \rho ^2 }}\left( {\left( {r^2  + a^2 } \right)^2  - a^2 \bar \Delta \left( {1 - x^2 } \right)} \right){\rm{d}}\phi ^2 \,,
	\ee 
	where
	\be 
	\bar \rho ^2  = r^2  + a^2 x^2 \,,
	\ee 
	and
	\be 
	\bar \Delta  = r^2  - 2Mr + a^2  + Q^2 \,.
	\ee
The accompanying gauge field is
	\be \label{gaugeKN}
	A_\mu  {\rm{d}}x^\mu   = \frac{{Qr}}{{\bar \rho ^2 }}\left( {{\rm{d}}t - a\left( {1 - x^2 } \right){\rm{d}}\phi } \right)\,.
	\ee 
Taking $Q\to 0$ limit, one recovers the Kerr solution.

\end{document}